\pdfoutput=1
\documentclass{article}
\usepackage{arxiv}
\usepackage{fvextra}  % extends fancyvrb, includes Verbatim
\DefineVerbatimEnvironment{jsoncode}{Verbatim}
  {fontsize=\scriptsize,
   breaklines,
   breakanywhere}
 \usepackage{float}
\usepackage[utf8]{inputenc} % allow utf-8 input
\usepackage[T1]{fontenc}    % use 8-bit T1 fonts
\usepackage{url}            % simple URL typesetting
\usepackage{booktabs}       % professional-quality tables
\usepackage{amsfonts}       % blackboard math symbols
\usepackage{nicefrac}       % compact symbols for 1/2, etc.
\floatstyle{ruled}
\newfloat{listing}{htbp}{lop}
\floatname{listing}{Listing}
\usepackage{inconsolata}  
\usepackage{microtype}      % microtypography
\usepackage{lipsum}		% Can be removed after putting your text content
\usepackage{graphicx}
\usepackage[colorlinks=true,linkcolor=blue,citecolor=blue,urlcolor=blue]{hyperref}
\usepackage{gensymb}
\usepackage{amsmath}

\title{MultiChain Blockchain Data Provenance for Deterministic Stream Processing with Kafka Streams: A Weather Data Case Study}

\author{ \href{https://orcid.org/0009-0008-0515-5462}{\includegraphics[scale=0.06]{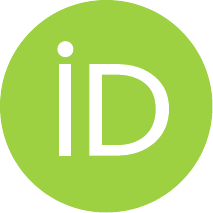}\hspace{1mm}Niaz Mohammad Ramaki}\\
	BIFOLD \\
	Technicshe Universität Berlin\\
	Berlin, Germany \\
	\texttt{ramaki@tu-berlin.de} \\
	\And
	\href{https://orcid.org/0000-0003-4548-788X}{\includegraphics[scale=0.06]{figures/orcid.pdf}\hspace{1mm}Florian Schintke} \\
	Department of Distributed Algorithms\\
	Zuse Institute Berlin\\
	Berlin, Germany \\
	\texttt{schintke@zib.de} \\
	}

\renewcommand{\shorttitle}{\textit{arXiv} Template}

\begin{document}
\maketitle

\begin{abstract}
Auditability and reproducibility still are critical challenges for real-time data streams pipelines. Streaming engines are highly dependent on runtime scheduling, window triggers, arrival orders, and uncertainties such as network jitters. These all derive the streaming pipeline platforms to throw non-determinist outputs. In this work, we introduce a blockchain-backed provenance architecture for streaming platform (e.g Kafka Streams) the publishes cryptographic data of a windowed data stream without publishing window payloads on-chain. We used real-time weather data from weather stations in Berlin. Weather records are canonicalized, deduplicated, and aggregated per window, then serialised deterministically. Furthermore, the Merkle root of the records within the window is computed and stored alongside with Kafka offsets boundaries to MultiChain blockchain streams as checkpoints.
Our design can enable an independent auditor to verify:
(1) the completeness of window payloads, 
(2) canonical serialization,  and 
(3) correctness of derived analytics such as minimum/maximum/average temperatures.
We evaluated our system using real data stream from two weather stations (Berlin-Brandenburg and Berlin-Tempelhof) and showed linear verification cost, deterministic reproducibility, and with a scalable off-chain storage with on-chain cryptographic anchoring.
We also demonstrated that the blockchain can afford to be integrated with streaming platforms particularly with our system, and we get satisfactory transactions per second values.

\end{abstract}

% keywords can be removed

\section{Introduction}
\label{lbl:introduction}
%%%%%%%%%%%%%%%%%%%%%%%%%%%%%%%%%%%%%%%%%%%%%%%%%%%%%%%%%%%%%
Modern stream processing platforms such as Kafka Streams, Apache Flink have the ability to handle continuous data streams by performing real-time computations, enabling low latency processing like filtering, joining, aggregating, of data while it arrives~\cite{carbone_apache_nodate}. 
Stream processing platforms usually receive data streams from distributed streams sources such as IoT sensors, or output of other stream processing systems. Stream processing engines also reduce the amount of data to be kept for further processing (e.g. extracting relative fields from data records). The most significant duties of stream processing platforms are to perform record processing with low latency, real-time, without needing costly storage~\cite{stonebraker2005}. These characteristics of stream processing platforms on the other hand require more observation in case of testing accountability and trust. Auditability and reproducibility turn to become harder when there are sophisticated streaming operations (e.g. window joins) running on a streaming pipeline~\cite{pecj2024}.
 Usually, when  stream results are produced, the outputs are forwarded down stream but without providing any footprints to be backward traceable.  This feature makes auditability tasks for stream processing engines a bit challenging.  We have already discussed the reasons which cause challenges to auditibility and reproducibility in streaming platforms in our previous work in~\cite{ramaki2025}. The main reason for aduditibility originate from non-deterministic nature on stream processing platforms due several internal and external factors, which turns the processing results rarely audible and reproducible. 

\begin{figure}[H]
\centering
\includegraphics[width=1\linewidth]{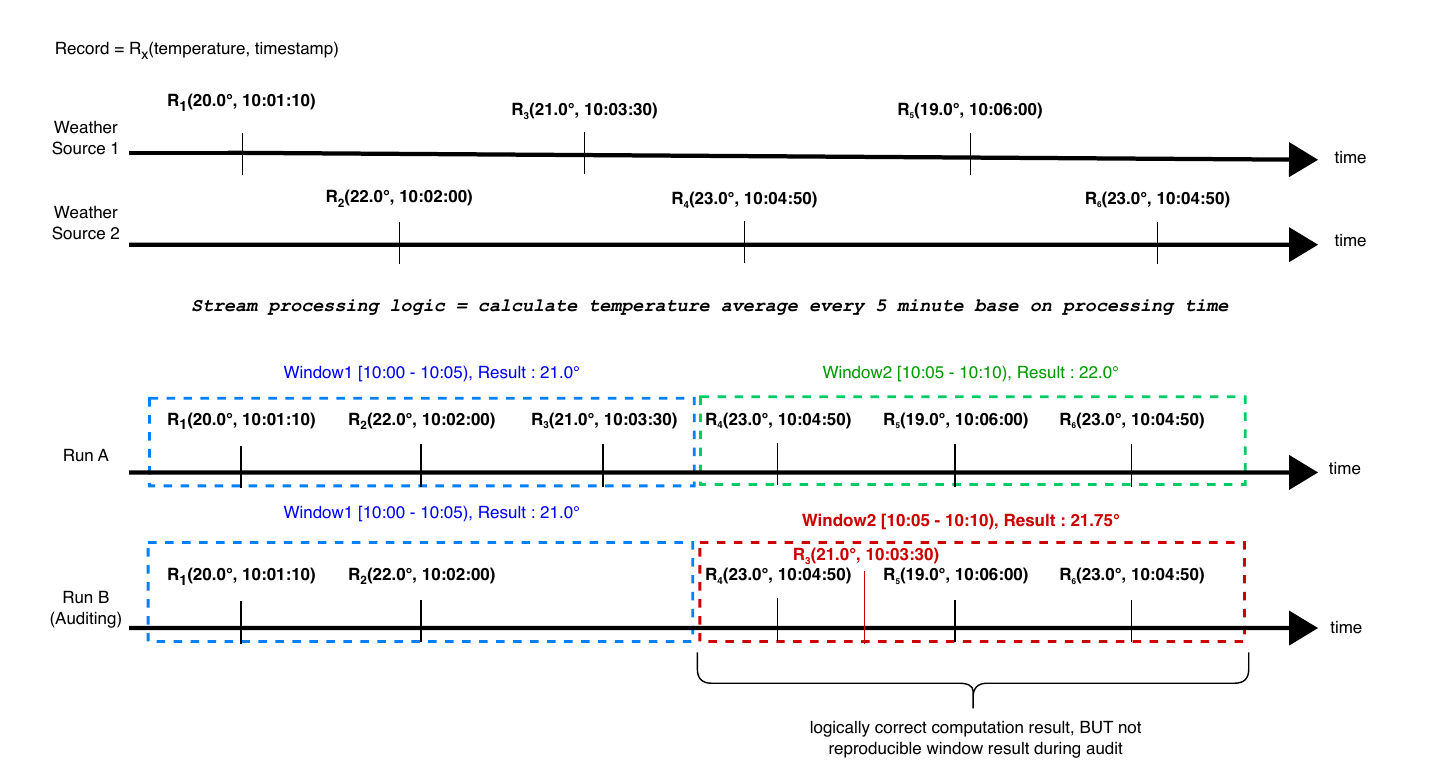}
\caption{Two weather stations emitting a regional temperature to a stream processing pipeline. The processing logic builds stream window base on processing time $\leq$ ingestion time. Streaming pipeline produces two different results in each execution.}
\label{fig:scenario}
\end{figure}

\textbf{An execution scenario:} Let's consider two weather stations across a city. Referring to figure~\ref{fig:scenario}, each weather source emits temperature records to a streaming pipeline. The streaming logic computes the average of the temperature every 5 minutes, it constructs the windows based on the processing time. Populating records into windows are non-deterministic, meaning window membership is not decided according to the timestamps which each record carries(i.e. event-time), but when the stream processor happens to read a record. At first execution, the processor fits three records into each window, here two windows, and calculates the average and emits the result into downstream (window1 = 21.0\degree and window2 = 22.0\degree). Later for the sake of audit or reproducibility, the second execution is performed, this time the record $R_{3}$ arrives due to external factors such as network jitter or internal factors such as rebalancing. This time $R_{3}$ fits into window2, and subsequently the result is not identical to the first run.
Current blockchain-backed provenance approaches mostly use hash chains to store data fingerprints. 
 Hash-chains bind provenance to ingestion order and cannot be replayed deterministically under re-keyed windows~\cite{bstprov2022}.
Although hash chains are favourable strategy for data provenance, but by design they come up with two major drawbacks, (1) storage overheads, ---as long as the chain gets bigger, it need careful management and more storage, (2) computation cost, --- repeated hashing for very large chains during verification. Taking these demerits into observation, we propose Merkle-tree based window provenance. 
Rather than hashing individual events in sequence, we cryptographically seal every window of data using canonical JSON~\cite{json2020} and a deterministic Merkle tree~\cite{cmt2025,smt2016}.
The payload remains off-chain; the commitment and metadata are immutably stored in a blockchain stream.
An auditor can later reconstruct the complete window deterministically.
The rest of this text contains the following sections:\\
In sections~\ref{lbl:streaming-platform}, and~\ref{lbl:multichain}, we introduced Kafka Streams platform and the MultiChain blockchain respectively. 
Section~\ref{lbl:audit} discusses formally the stream auditability, section~\ref{lbl:implementation} describes our systems architecture, section~\ref{lbl:verification} explains our verification methods, section~\ref{lbl:evaluation} is about our evaluation methods and approaches, section~\ref{lbl:related-work} narrates the related work, section~\ref{lbl:discussion} is the discussion part, and section~\ref{lbl:conclusion} is the conclusion of the paper.  

\vspace{2mm}

%%%%%%%%%%%%%%%%%%%%%%%%%%%%%%%%%%%%%%%%%%%%%%%%%%%%%%%%%%%%%%%%
\section{Streaming Platform}
\label{lbl:streaming-platform}
\begin{figure}[H]
\centering
\includegraphics[width=0.9\linewidth]{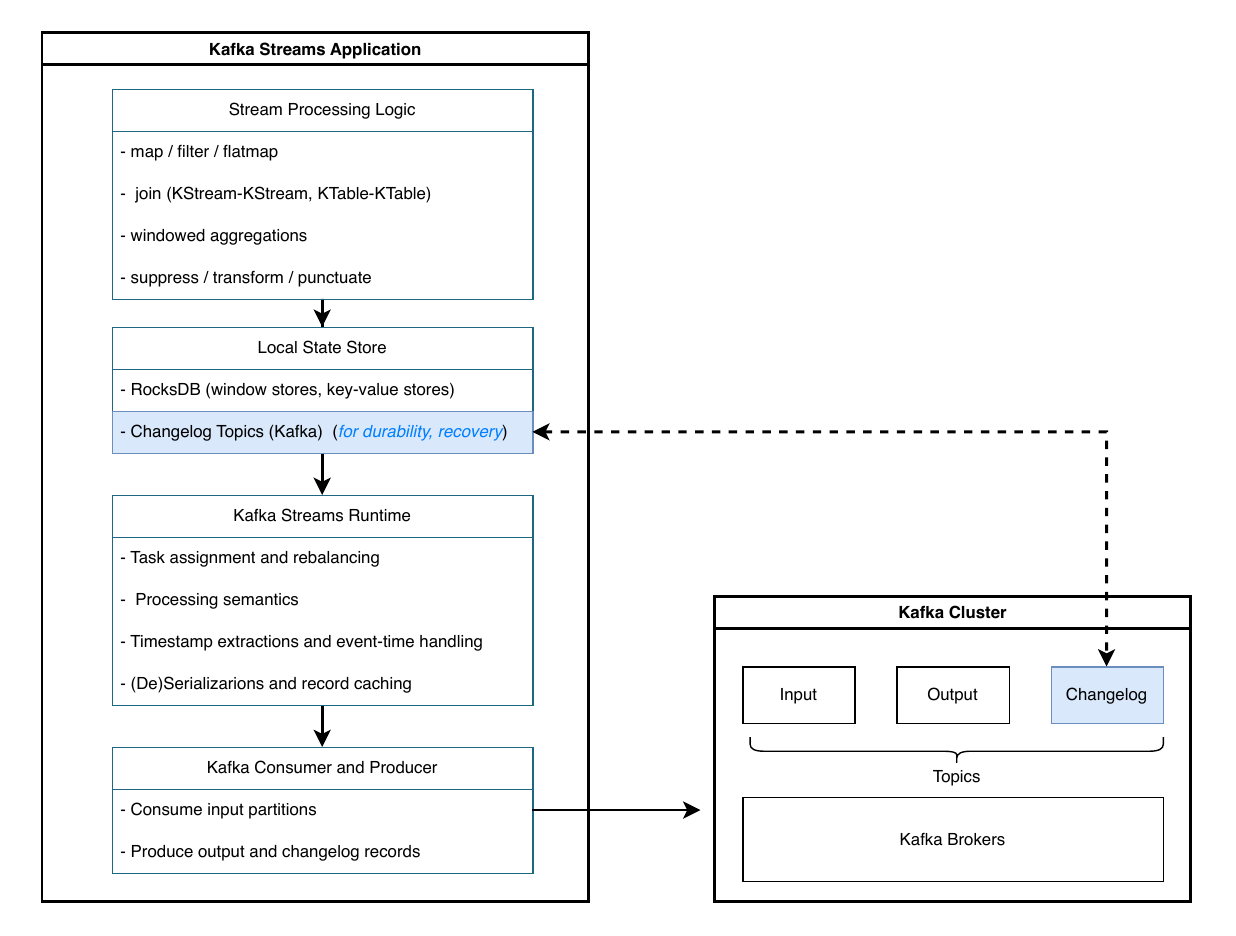}
\caption{Kafka Streams Architecture}
\label{fig:kafka-streams}
\end{figure}

A streaming platforms, for example Kafka Stream~\cite{kafka-streams}, is designed to process real-time unbounded data continuously as it arrives. It can transform, aggregate, and correlate events with low latencies and strong scalability guarantees. Kafka Streams is a java library which directly runs inside a JVM application, consumes and produces data to Kafka topics. Kafka Streams uses Kafka for durability, partitioning, and fault tolerance. For a stream processing logic, it provides high-level declarative operators in a DSL package. It also supports low level processing APIs for defining complex processing logic using Processor API. For stateful stream processing, it maintains a local state store database(i.e. RocksDB), which is backed by \emph{changelong} topics for fault tolerance. Kafka Streams supports three processing semantics (at-least once, at-most once, and exactly once). Figure~\ref{fig:kafka-streams} shows the architecture of Kafka Streams processing platform.

\vspace{2mm}

%%%%%%%%%%%%%%%%%%%%%%%%%%%%%%%%%%%%%%%%%%%%%%%%%%%%%%%%%%%%%%%%
\section{MultiChain Blockchain}
\label{lbl:multichain}
\begin{figure}[H]
\centering
\includegraphics[width=0.8\linewidth]{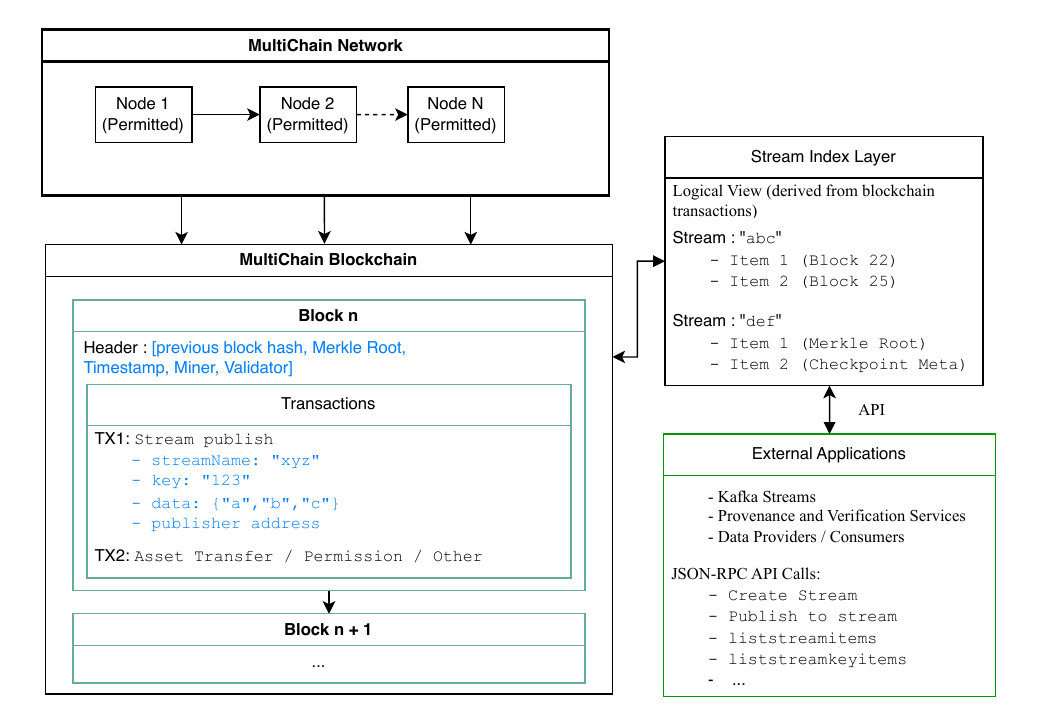}
\caption{MultiChain Architecture}
\label{fig:multichain}
\end{figure}

MulltiChain~\cite{multichain} is a permissioned blockchain for enterprises which need controlled, auditable and high performance distributed ledger instead of an open cryptocurrency network. It uses Bitcoin core but replaces proof-of-work with round-robin mining model. MultiChain has fine-grained permissions (connect, send, receive, mine, admin, issue, activate) to control nodes in the blockchain. It intentionally avoids smart contracts to reduce complexity and nondeterminism on the chain in favour of external application logic. Applications can interact with MultiChain through their already provided JSON-RPC APIs.  MultiChains features an append-only, key-value, and time-ordered database called \emph{streams}, which are suitable for event-logs and provenance records. Streams are papular for data anchoring, and audit trails rather than smart contracts. Streams are created explicitly and data are stored to them through  applications using API commands~\cite{multichain-apis}.  Streams support a clean off-chain/on-chain split. Off-chain files are large mutable data which stay out side of blockchain, while streams in contrast store small deterministic anchors (hashes, metadata), enabling reproducibility and auditability without letting blockchains to bloat by storing large data.  Figure~\ref{fig:multichain} illustrates architecture of a MultiChain blockchain emphasizing in stream features. 
\vspace{2mm}

\section{Stream Auditability}
\label{lbl:audit}
A stream processing system is auditable if the system keeps the lineage of the process of an input until the output is produced, such that an output result \emph{y} can be traced back to the followings:
\begin{itemize}
	\item[-] the input set $X_{k,w}$, ---where there is a key \emph{k} for a window \emph{w}
	\item[-] the execution environment $\rho$
	\item[-] and verifiable anchor $\alpha$ (e.g. Merke root, blockchain TxID, or cryptographic hash) that links the output \emph{y} to its origin. 
\end{itemize} 
\begin{equation}
audit(y) = (X_{k,w}(F,C), \rho, \alpha)
\end{equation}
\begin{itemize}
	\item[-] $\alpha\in$H($X_{k,w}$), where H is Merkle root or a secure hash, and stored in an immutable verifiable medium such as blockchain.
\end{itemize}

%%%%%%%%%%%%%%%%%%%%%%%%%%%%%%%%%%%%%%%%%%%%%%%%%%%%%%%%%%%%%
\section{Implementation}
\label{lbl:implementation}
To implement our proposed solution for a stream window provenance, we use realtime open access data from German Metrological Service (DWD), provided by BrighSky API\footnote{https://brightsky.dev}. Our Kafka source connector~\footnote{\url{https://github.com/niazmr/MyApiSourceConnector}} pulls weather data from two weather stations from two different regions in Berlin, named "Berlin-Brandenburg" and "Berlin-Tempelhof". Our processing pipeline is not complex. First the two streams are merged. Then it computes region based  and Berlin based a window-based statistics (such as avg, min, max) within a specific window and emits these results to down stream.
\subsection{System Architecture}
\label{sect:arch}
%%%%%%%%%%%%%%%%%%%%%%%%%%%%%%%%%%%%%%%%%%%%%%%%%%%%%%%%%%%%%
Figure~\ref{fig:arch} illustrates our architecture.
The system comprises three components:
\begin{enumerate}
    \item \textbf{Ingestion.}
    Stream of weather records coming from "Berlin-Brandenburg" and "Berlin-Tempelhof" weather stations. The streams are injected into different Kafka topics.  Each record contains station identifier, timestamp, and temperature, and some other fields.
    \item \textbf{Processor.}
    Kafka Streams constructs fixed tumbling windows over weather station keys and performs aggregation of minimum, maximum, and average temperatures. Simultaneously, each incoming weather records is buffered by a Checkponiter component.
\item \textbf{Blockchain anchoring.}
When a window is closed, it is possible that the window may contain a multi-set of records, therefore the payload items are deduplicated and canonicalized. Then the Merkle root of that window is constructed and the following checkpoint data is stored on chain:

\begin{listing}[htbp]
  \caption{Checkpoints from weather sources “Berlin-Brandenburg” (left) and “Berlin-Tempelhof” (right) on the blockchain}
  \label{lst:checkpoints-bb-bt}
  \centering
  \begin{minipage}[t]{0.48\textwidth}
  \begin{jsoncode}[frame=single]
{
  "checkpoint": {
    "blockchainStream": "BrandenburgCheck",
    "merkleRoot": "9d1336c6308841e556058a2251bb495bc679ed050f53646ce21e200af35a991e",
    "offsetEnd": 9,
    "offsetStart": 6,
    "payloadPath": "Files/payloads/Berlin Brandenburg_2025-12-02T18_00_00Z_2025-12-02T20_00_00Z.json",
    "payloadSha256": "253d33d44a48f912085a1ec48c79ae5eb63087fad336c8d4f212d681f09d831c",
    "recordCount": 4,
    "sourceStream": "Berlin Brandenburg",
    "windowEnd": "2025-12-02T20:00:00Z",
    "windowId": "Berlin Brandenburg:2025-12-02T18:00:00Z_2025-12-02T20:00:00Z",
    "windowStart": "2025-12-02T18:00:00Z"
  }
}
  \end{jsoncode}
  \end{minipage}\hfill
  \begin{minipage}[t]{0.48\textwidth}
  \begin{jsoncode}[frame=single]
{
  "checkpoint": {
    "blockchainStream": "TempelhofCheck",
    "merkleRoot": "dba5f2f40511466834a67bcfe79e549681e9e5703dc147daf68b5d599690d63d",
    "offsetEnd": 5,
    "offsetStart": 5,
    "payloadPath": "Files/payloads/Berlin-Tempelhof_2025-12-02T16_00_00Z_2025-12-02T18_00_00Z.json",
    "payloadSha256": "ec41ffac7dd84f7118f447fefac4e5d893fd79de5f37da13e0455b8ec7815485",
    "recordCount": 1,
    "sourceStream": "Berlin-Tempelhof",
    "windowEnd": "2025-12-02T18:00:00Z",
    "windowId": "Berlin-Tempelhof:2025-12-02T16:00:00Z_2025-12-02T18:00:00Z",
    "windowStart": "2025-12-02T16:00:00Z"
  }
}
  \end{jsoncode}
  \end{minipage}
\end{listing}
\end{enumerate}
\vspace{2mm}
\begin{figure}[H]
\centering
\includegraphics[width=0.9\linewidth]{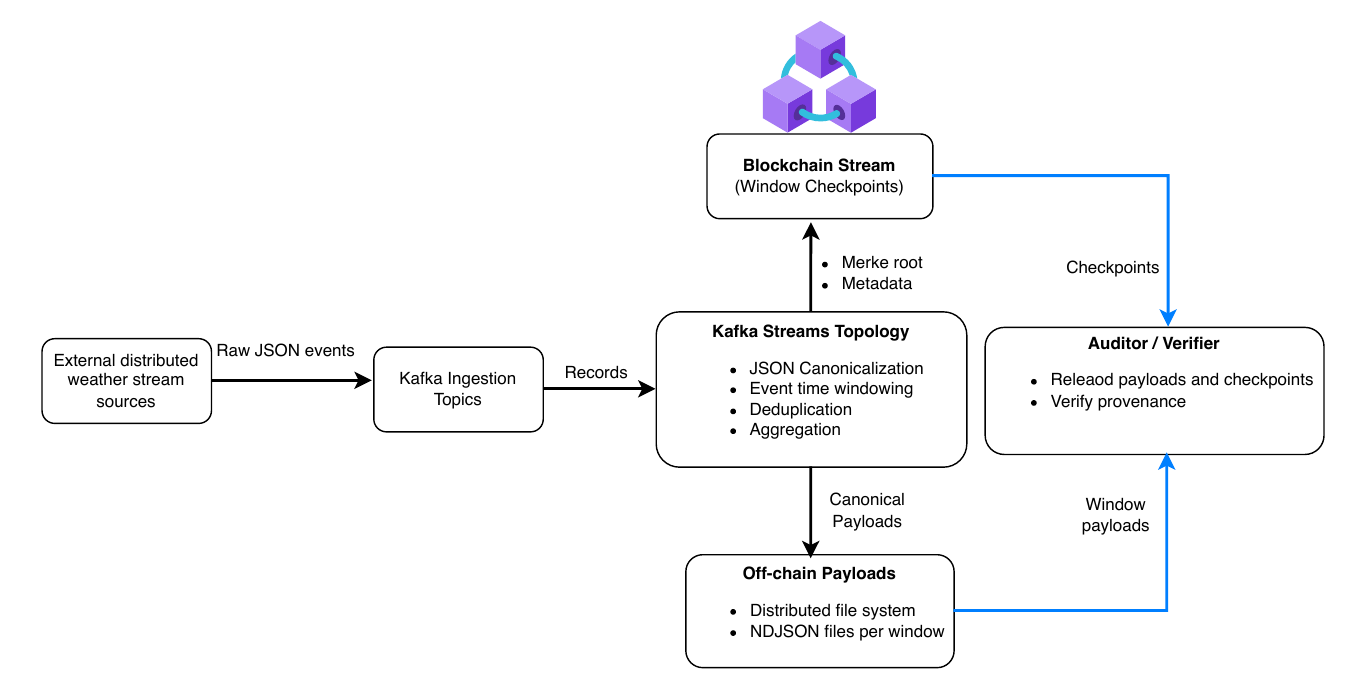}
\caption{
Architecture overview:
 Weather stream sources $\rightarrow$ merge weather sources $\rightarrow$ deterministic windowing $\rightarrow$ canonicalization and Merkle commitment $\rightarrow$ checkpoint anchoring $\rightarrow$ auditor verification.}
\label{fig:arch}
\end{figure}

%%%%%%%%%%%%%%%%%%%%%%%%%%%%%%%%%%%%%%%%%%%%%%%%%%%%%%%%%%%%%
\subsection{Deterministic Windowing}
%%%%%%%%%%%%%%%%%%%%%%%%%%%%%%%%%%%%%%%%%%%%%%%%%%%%%%%%%%%%%
We implemented epoch-aligned tumbling windows, which are fixed windows and their boundaries always starts at fixed absolute times, not relative to when the systems started. The reason here we use epoch-aligned tumbling windows is, because it is, deterministic, global and absolute. 
Our epoch-aligned tumbling windows  duration is $\Delta$, and  $\Delta$ = (windowEnd - windowStart) represented in seconds.
In the above Checkpoint examples, we have one tumbling window of size $\Delta$ = 2 hours

For an event with timestamp $t$, the window index is:
\begin{equation}
w = \left\lfloor \frac{t}{\Delta} \right\rfloor.
\end{equation}
\emph{t} = window start in seconds since epoch (Unix time). Here window index, \emph{w} is the integer number that uniquely identifies to which fixed window a timestamp of an event belongs.\\
The physical inverval of window  $w$ is:
\begin{equation}
I_{w} = [w\Delta,(w+1)\Delta).
\end{equation}
Let's call $I_{w}$ window identifier of window \emph{w}.  \\
All the records or events belonging to $I_{w}$ are going to be processed as one batch, which are independent of ingestion time, or watermark delays
\vspace{2mm}
\begin{figure}[H]
\centering
\includegraphics[width=0.75\linewidth]{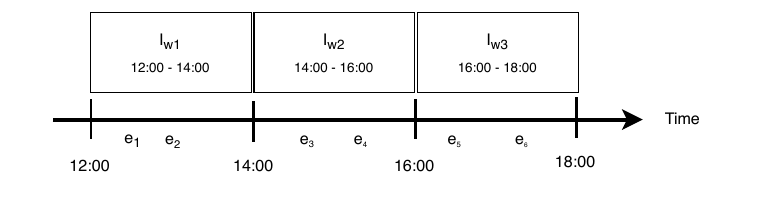}
\caption{Epoch-based windowing: events map deterministically to $I_w$ independent of processing order. All events with timestamps in $I_{w_{I}}$ are grouped into the same window.}
\label{fig:window}
\end{figure}

%%%%%%%%%%%%%%%%%%%%%%%%%%%%%%%%%%%%%%%%%%%%%%%%%%%%%%%%%%%%%
\subsection{Canonical JSON and Merkle Tree Construction}
%%%%%%%%%%%%%%%%%%%%%%%%%%%%%%%%%%%%%%%%%%%%%%%%%%%%%%%%%%%%%
Canonical JSON~\cite{json2020} makes certain that each weather record is represented in a deterministic, byte-uniform regardless where and how it was processed. Fields serlialization is stable and lexicographically ordered, timestamps are normalized to UTC with second level precision. In canonical JSON, no realtime dependency is allowed.  Canonicalization gets rid of  non-determinism from ingestion order, scheduling, or batching effects within Kafka Streams. After canonicalization, records deduplications is performed, then each record becomes a leaf of the Merkle tree. After computing a SHA-256 digest, all leaves are sorted lexicographically to enforce stable ordering. The final Merkle root becomes a compact integrity truth to the entire window. Listing~\ref{lst:mt} illustrates how Merkle root of a window is constructed. \\

\begin{figure}[H]
\centering
\includegraphics[width=0.6\linewidth]{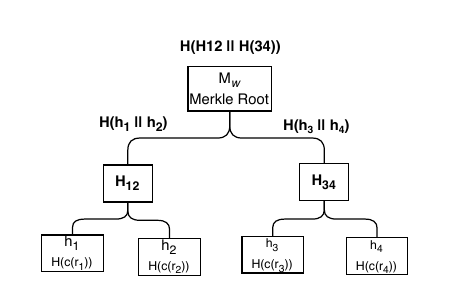}
\caption{Merkle tree of canonical payload. Rebuilding $M_w$ proves payload completeness. Canonical JSON strings $C(r_i)$ are hashed to leaves $h_i$. Internal nodes hash concatenations of their children. The root $M_w$ uniquely commits to the entire window payload.}
\label{fig:merkle}
\end{figure}

Below is the steps of JSON canonicalization of each Weather $r$ in a window:
\begin{enumerate}
\item remove fields which cause inconsistency,
\item convert timestamps to UTC ISO-8601,
\item lexicographically sort object keys.
\end{enumerate}
\textit{Constructing Merkle Root:} Let $C(r)$ be the canonical string.
We compute leaf hashes:
\begin{equation}
h_i = H(C(r_i)),
\end{equation}
with hashing,  $H = \text{SHA-256}$. The leaf nodes  should be  sorted lexicographically.
Merkle internal nodes compute the hash of concatenation of left and right wings of the tree.
\begin{equation}
H_{parent} = H(h_L \Vert h_R).
\end{equation}
The Merkle root $M_{w}$ in figure ~\ref{fig:merkle} cryptographically binds all records.
\vspace{1mm}
\begin{listing}[htbp]
  \caption{Merkle root calculation pseudo code}
  \label{lst:mt}
  \centering
  \begin{jsoncode}
merle_root(values):
    if values is empty:
        return NULL
    // leaf hashes
    leaves <-- []
    for each v in values do
        leaves.append(SHA256(v))
    sort(leaves)                       // lexicographic order
    level <-- leaves
    while lenght(level) > 1 do
        nextLevel <-- []
        i <-- 0
        while i < length(level) do
            left  <-- level[i]
            if i+1 < length(level) then
                right <-- level[i+1]
            else
                right <-- left           // duplicate last leaf if odd
            nextLevel.append(SHA256(left || right))
            i <-- i + 2
        level <-- nextLevel
    return level[0]                    // Merkle root
  \end{jsoncode}
\end{listing}

%%%%%%%%%%%%%%%%%%%%%%%%%%%%%%%%%%%%%%%%%%%%%%%%%%%%%%%%
\subsection{Blockchain Checkpointing}
%%%%%%%%%%%%%%%%%%%%%%%%%%%%%%%%%%%%%%%%%%%%%%%%%%%%%%%%%%%%%
As mentioned earlier in section~\ref{sect:arch}, we store checkpoint data on specific data streams on MultiChain~\cite{multichain} blockchain. For each weather source, we have a data stream on the blockchain which stores the checkpoint data of that weather source stream. 
Important data items of a checkpoint data include the followings:
\begin{itemize}
\item[-] $\texttt{sourceStream},$ weather station ID,
\item[-] $\texttt{windowStart}$ and $\texttt{windowEnd}$,
\item[-] $\texttt{recordCount},$ the number records within a window,
\item[-] $\texttt{payload path},$ path where the off-chain payloads are stored. This can be a distributed files system too.
\item[-] $\texttt{payloadHash},$ the hash of the payload ensures the integrity of the payload
\item[-] $\texttt{offsetStart}$ and $\texttt{offsetEnd}$, these are offsets of the Kafka partitions 
\item[-] $\texttt{merkleRoot}$ of the window payloads,
\item[-] $\texttt{windowID}$, e.g.  "Berlin Brandenburg:2025-12-02T18:00:00Z\_2025-12-02T20:00:00Z".
\end{itemize}
We also log the same checkpoint data off-chain, but annotating each checkpoint data with its corresponding transactions ID from the blockchain.

The actual data payloads remain off-chain (e.g. distributed filesystem, database, or local disk) under append-only NDJSON~\cite{ndjson}. The reason we use NDJSON instead of a JSON array are the following reasons:
\begin{itemize}
	\item[-] \emph{Memory Efficiency:} JSON needs to load the whole array, while NDJSON processes one line at a time, $\rightarrow$ saving memory,
	\item[-] \emph{Streaming:} NDJSON is stream-friendly, $\rightarrow$ perfect for realtime data,
	\item[-] \emph{Parallelism:} NDJSN easily split and processed by multiple workers,
	\item[-] \emph{Simplicity:} Each line in NDSON file is a valid JSON object, making it simple to parse individual records. 
\end{itemize}

\section{Verification}
\label{lbl:verification}
Verification of windows are independent of Kafka it enables cross-organizational reproducibility. 
An auditor performs verification with the algorithm listed in Listing~\ref{lst:verification} for validating a window \emph{W} provenance. Auditors can independently reconstruct cryptographic commitment of a window such as $\hat{R}$ recomputed Merkle root, and  $\hat{N}$ recomputed number of records from a checkpoint data, and compare with their corresponding  metadata (i.e. M=merkle root, N=record count) of checkpoints from blockchain. A window provenance verification is passed if only if the following condition is met:
\begin{equation}
\text{Verified}(W) = (\hat{R} = R) \;\land\; (\hat{N} = N).
\end{equation}  

\begin{listing}[htbp]
  \caption{Window provenance verification pseudo code}
  \label{lst:verification}
  \centering
\begin{jsoncode}
Input:
    windowId        // e.g. "Berlin-Tempelhof:2025-12-02T14:00:00Z_2025-12-02T16:00:00Z"
    checkpoints_data  //  blockchain API
Output:
    Verified or Failed
function verify_window(windowId, checkpoints_data):
    // 1. Load checkpoint metadata
    checkpoint <-- find_checkpoint(checkpointsLog, windowId)
    if checkpoint_not_found then
        return Failed  				    // no evidence for this window
    R      <-- checkpoint.merkleRoot
    N      <-- checkpoint.recordCount
    path   <-- checkpoint.payloadPath      // e.g. "Files/payloads/<...>.json"
    // 2. Load off-chain payload
    payloadRecords <-- load_Json_objects(path)
    if payloadRecords_is_empty or error_in_payloadRecords then
        return Failed
    // 3. Canonicalize all records and count them
    canonicalList <-- empty list
    for each record r in payloadRecords do
        c <-- canonical_Json(r)           		// deterministic serialization
        append c to canonicalList
    N_hat <-- length(canonicalList)
    // 4. Rebuild Merkle tree
    R_hat <-- merkle_root(canonicalList)  		// sorted leaves + SHA-256 tree
    // 5. Compare
    if (R_hat = R) and (N_hat = N) then
        return Verified
    else
        return Failed
\end{jsoncode}
\end{listing}
\begin{figure}[H]
\centering
\includegraphics[width=0.6\linewidth]{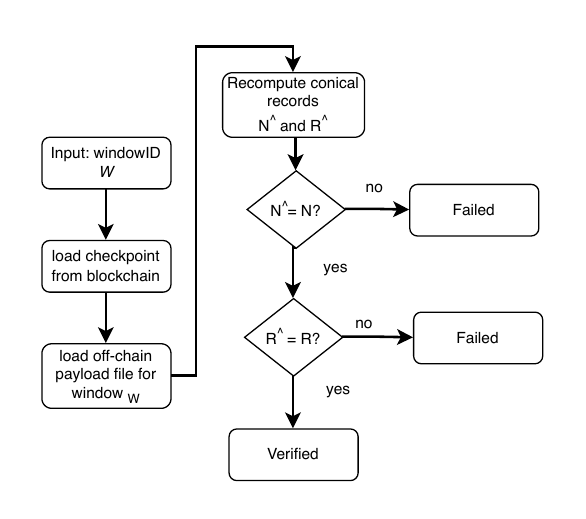}
\caption{Verification decision procedure for window $W$. The window is accepted as intact if and only if both record count and Merkle root match the checkpoint.}
\label{fig:arch}
\end{figure}

\subsection{Reproducibility}
We performed independent runs of ingestion in two machines:
\begin{itemize}
	\item[-] MacOS M2, Java 21
	\item[-] MacOS M4, Java 21
\end{itemize}
Both produced byte identical results:
\begin{itemize}
	\item[-] canonical JSON
	\item[-] payload SHA-256
	\item[-] Merke roots
\end{itemize} 
Which confirms canonicalization eliminates environment-induced nondeterminism. 
%%%%%%%%%%%%%%%%%%%%%%%%%%%%%%%%%%%%%%%%%%%%%%%%%%%%%%%%%%%%%
\section{Evaluation}
\label{lbl:evaluation}
%%%%%%%%%%%%%%%%%%%%%%%%%%%%%%%%%%%%%%%%%%%%%%%%%%%%%%%%%%%%%

We ingested sensor data streams from Berlin-Tempelhof and Berlin-Brandenburg weather stations. Each produced window is anchored to MultiChain.

\subsection{Merkle Verification Cost}
\begin{figure}[H]
\centering
\includegraphics[width=1\linewidth]{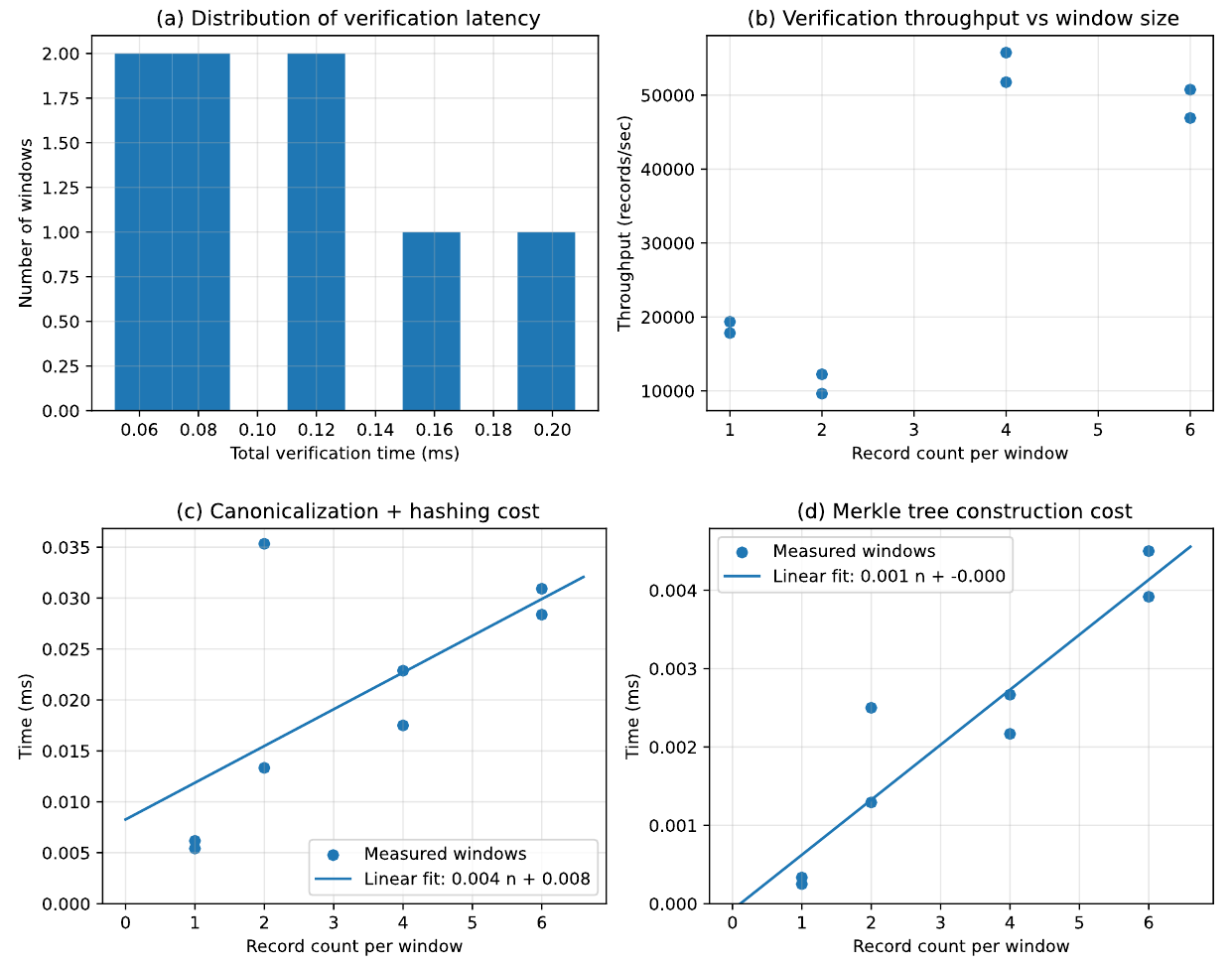}
\caption{(a) showing relationship between record count and end-to-end verification latency. (b) illustrates that verification process is stable across all evaluated windows. (c) demonstrates high throughput even for windows with larger number of records. (d) Merkle tree construction remains near-linear scaling. Together, the results ensure that cryptographic provenance verification can be carried out efficiently with negligible overhead, and is acceptable for frequent and on-demand demand auditing in real-time streaming systems.}
\label{fig:verification}
\end{figure}

Let $T(n)$ be end-to-end verification time.
Empirically:
\[
T(n) = \alpha n + \beta,
\]
Where \emph{n} is the number of records, $\alpha$ capturing hashing cost and $\beta$ fixed I/O expense. Observed $\alpha$ is within linear bounds, consistent with tree depth $O(\log n)$ hashed in batches. The slope $\alpha$ corresponds to marginal cost per additional records, which shows how many seconds or microseconds are added to it when one more record is added to the window. The parameter $\beta$ is independent of the number of records, it only captures the fixed overheads such as I/O and object initialization. \\
In figure~\ref{fig:verification} plot (a) shows the distribution of end-to-end  verification latency across all evaluated windows. The distributions seem to be narrow and compact, indicating stable, predictable and consistent latency across all windows. There are no long-tail outliers, which shows on modern hardware canonicalization, hashing, and Merkle reconstruction variance is negligible. The second plot (b), showcase the system throughput during verification (records/second), which is extremely high. Each dot represents one verified window. Throughput is computed according to:
\[
throughput =  \frac{recordCount}{verificationTime} 
\]
From this plot we guess, when the number of records in a window (i.e. window size) grows, the throughput does not collapse.
The plot (c) in figure~\ref{fig:verification} shows the cost of converting each record into canonical JSON and hashing them using SHA-256.  The regression line is not for any prediction, instead it is used to quantify the scaling behaviour of canonicalization and hashing. Each dot represents a weather record. The outlier dot shows only a runtime fluctuation not a semantic anomaly. Overall, the linear regression fitted through all dots, indicating when the number of records within a window increases, the canonicalization and hashing increases proportionally. Likewise,  plot (d) in the same figure, measures the cost of constructing the Merkle tree from the sorted leaf hashes. 

\subsection{Blockchain}
\begin{figure}[H]
\centering
\includegraphics[width=1\linewidth]{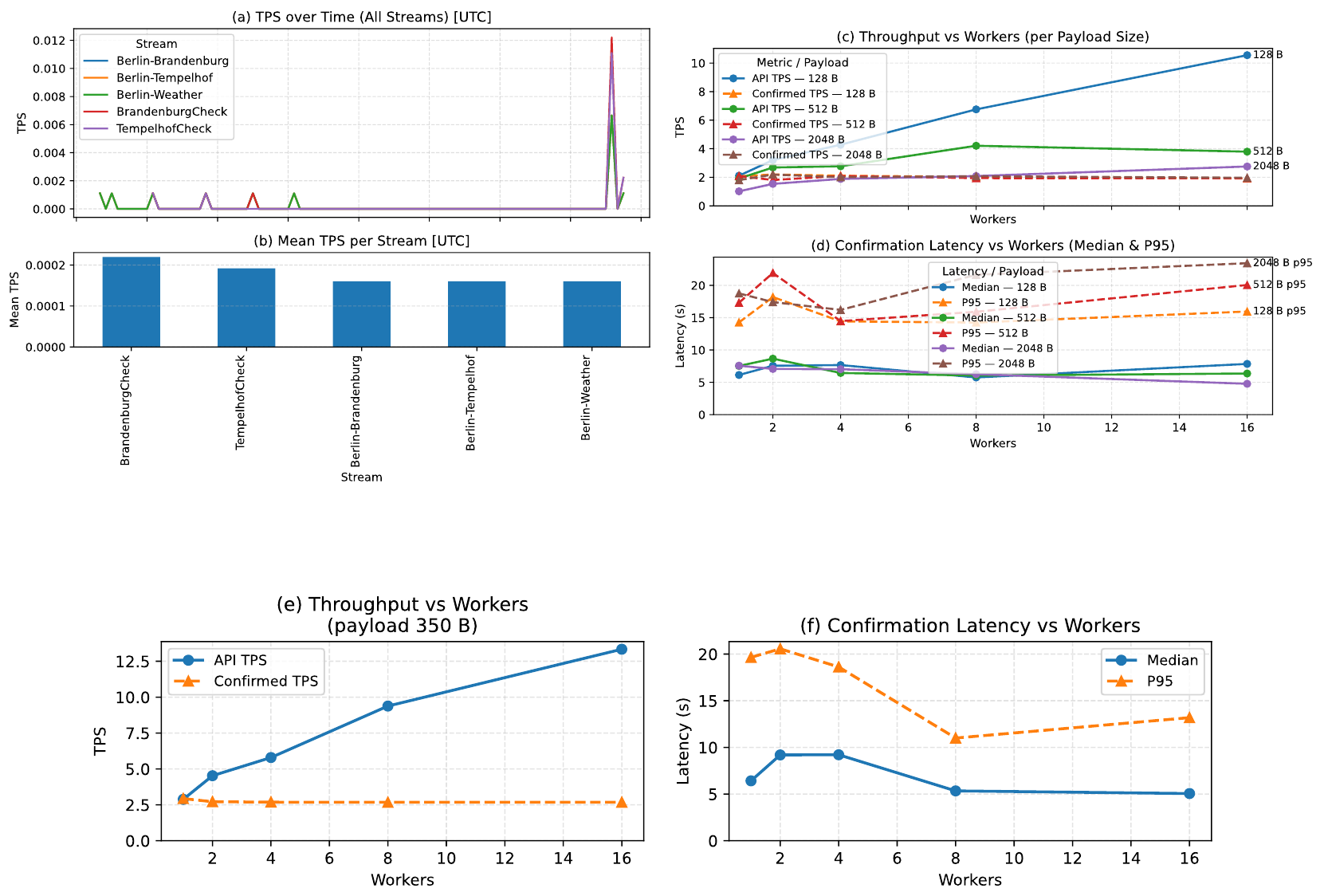}
\caption{a) Observed on-chain throughput (items/second) for the five blockchain streams. Shows 5-minute window throughput aggregation. The values (0–0.015 items/s) reflect the real-world ingestion rate of payloads to the streams, providing the baseline operational load for the blockchain system. (b) indicates average of TPS from each stream. (c) shows scalability of confirmed and API transaction throughput as a function of the number of concurrent publishers and payload size (128 Bytes, 512 Bytes, 2048 Bytes). Smaller payloads achieve higher throughput (up to approximately 2.1 tx/s), while larger payloads reduce throughput due to increased serialization and validation overhead. (c) demonstrate the average confirmation latency of parallel publishers. In (d), we experience bottleneck when we increase the number of publishers.  All configurations exhibit a common saturation region beyond eight workers, revealing the blockchain’s inherent consensus-rate bound.}
\label{fig:blockchain}
\end{figure}

To demonstrate how the MultiChain blockchain with default parameters~\cite{mc-params} adapts with our data stream pipeline, we retrieved all stream items from the current streams from the blockchain to investigate the transaction per second (TPS) parameter. This is illustrated in figure~\ref{fig:blockchain} plot (a) and (b), we expected low TPS values, because stream checkpointing occurs infrequently compared to the raw event rates. The TPS computed as:
\[
TPS = \frac{totalNumberOfConfirmedTXs}{ConfrimationTimeOfLastTX - ConfrimationTimeOfFirstTX}
\]
In our normal situation, figure~\ref{fig:blockchain} plot (a), the denominator is relatively a large number, which result	low TPS values.
Therefore, we have performed benchmarks by specifying different data payloads and workers and published on a stream on the blockchain. This time we received optimistic TPS values, which are illustrated in plots (c) and (d).  Furthermore, to investigate our API throughput, we published a particular payload (i.e. 350 Bytes) by different workers. This time we experienced more throughput when the number of workers increases, but the TPS remains stable influenced by the consensus algorithm, which is round-robin in MultiChain blockchain by default. This is illustrated in plot (e) and (d).

%%%%%%%%%%%%%%%%%%%%%%%%%%%%%%%%%%%%%%%%%%%%%%%%%%%%%%%%%%%%%
\section{Related Work}
\label{lbl:related-work}
%%%%%%%%%%%%%%%%%%%%%%%%%%%%%%%%%%%%%%%%%%%%%%%%%%%%%%%%%%%%%
For data provenance significant number of literature are available, but particularly for data stream provenance there are a few. We point out here to the most related work pertaining to our work. Palyvos-Giannas et al. in~\cite{genealog2018} introduce \emph{GeneaLog} for managing fine-grained data provenance in streaming systems, particularly focusing on edge computing environments. It examines how GeneaLog captures the origin of data and its transformations throughout the processing pipeline, enabling better data accountability, debugging, and understanding of data flows by implementing of streaming operators that allow for the tracking of data provenance.
Wiemers~\cite{wiemers2023} in his Masters Thesis discusses the benchmarking of checkpointing algorithms in stream processing engines. It examines the need for fast and consistent fault recovery in data streams, categorizing rollback recovery mechanisms into uncoordinated, coordinated, and communication-induced protocols. The document also evaluates these checkpointing algorithms across various metrics such as latency, throughput, recovery times, and network overhead, concluding that coordinated approaches generally outperform uncoordinated solutions. It guarantees exactly-once recovery but does not provide cryptographic evidence to third parties evaluating the system. Farhan et al. proposes a conceptual framework, VeriTrust, they emphasize the need for a robust governance model and highlights limitations of the study such as the absence of a full prototype and the necessity for empirical validation of performance benchmarks like latency and availability. The integration of zero-knowledge proofs is noted as a potential future enhancement, which indicates a focus on privacy-preserving technologies. They also propose the use of blockchain technology for establishing credibility in information dissemination. It presents a variety of research on leveraging artificial intelligence and machine learning for combating misinformation, and discusses the importance of accreditation, audits, and revocation operations for maintaining the integrity of digital identities and transactions.~\cite{farhan2025}. 
Han et al. propose a novel store-verify-and-forward architecture~\cite{han2026}, which improves the efficiency of data verification and retrieval. By utilizing a unique combination of data chunking, Merkle trees, and interval trees, their verification method enhances the data integrity assurance process in distributed storage systems. The authors emphasize the limitations of traditional methods that fail to efficiently verify data integrity, leading to vulnerabilities within these systems. They present an innovative approach that constructs a Merkle hash tree for data blocks, improving the speed and reliability of integrity checks. Furthermore, they introduce a transmission credential mechanism to secure data during transfers, ensuring that every node’s involvement is logged and verifiable.This work underlines the importance of real-time monitoring and auditing in maintaining data authenticity and integrity, addressing the pressing need for robust solutions in blockchain-based storage systems.
In general, the above literatures manifest great works in the realm of data integrity and privacy, but they seldom explored the field of data stream integrity and auditability.

\section{Discussion}
\label{lbl:discussion}
In this work, we have used cryptographic tools such as hashing and Merkle root with the help of MultiChain blockchain to verify end-to-end provenance of a stream processing result. Predominantly concerning data provenance Merkle root and hash chains the are the mostly two used techniques. The two guarantee \emph{integrity}, and don't explain any \emph{meaning}. If they both \emph{purely} used in the context of a simple stream processing pipeline, they verify end-to-end provenance, meaning  tracing a result directly to the input of that result. In the case of complex streaming pipelines (e.g. window-joins), we need granularity of provenance steps to observe how step by step a result is concluded. In that case, Merkle root alone does not achieve the goal. Therefore, to achieve that goal a semantic layer should be added with this process, such as RDF~\cite{rdf}. \\
Moreover since verification cost of hash chains is linear, it will be not a proper solution for large number of records within a window. Although it is a satisfiable solution for less number of records. Therefore, Merkle root with logarithmic verification cost is an optimal solution.

%%%%%%%%%%%%%%%%%%%%%%%%%%%%%%%%%%%%%%%%%%%%%%%%%%%%%%%%%%%%%
\section{Conclusion and Future works}
\label{lbl:conclusion}
%%%%%%%%%%%%%%%%%%%%%%%%%%%%%%%%%%%%%%%%%%%%%%%%%%%%%%%%%%%%%
 Stream processing results verification still remains an open challenge. In this work, we introduced blockchain technology as an external, immutable reference layer for making stream processing results verifiable and auditable. Blockchain provides data provenance in streaming pipeline. Auditors can independently verify data streams computation from blockchain stored metadata and off-chain data. An auditor can supply a window ID information to the system and gets a verification message pertaining to that specific window. Data stream of a window are canonically serialized and the Merkle root of the records within that window is computed and published on the blockchain.  Our design scale to continues data ingestion and provides deterministic reprocessing under real operation conditions.
For our future task, we aim to include more programability to the blockchain-side using smart contracts or chaincodes to develop more granularity for step by step provenance tracking. Moreover adding semantics information for complex streaming pipelines can also be our next goal. Including zero-knowledge proofs is also a choice when data stream confidentiality matters and we don't want them to be exposed to the auditors, particularly when there is multi-party stream verifiers.
\bibliographystyle{plainnat}
\bibliography{bibliography.bib}  %%% Uncomment this line and comment out the ``thebibliography'' section below to use the external .bib file (using bibtex) .





 %\end{thebibliography}

\end{document}